\documentclass[aps,reprint,groupedaddress,longbibliography]{revtex4-1}

\usepackage{hyperref}
\hypersetup{
  colorlinks   = true, %Colours links instead of ugly boxes
  urlcolor     = blue, %Colour for external hyperlinks
  linkcolor    = blue, %Colour of internal links
  citecolor   = blue %Colour of citations
}
\usepackage{amsmath}
\usepackage{amssymb}

\usepackage[caption=false]{subfig}
\usepackage{tikz}
\usepackage{xifthen}
\usepackage{tikz-cd}
\usetikzlibrary{decorations.pathmorphing,calc,patterns,intersections,
calc,angles,quotes,shapes,arrows,decorations.markings,matrix, arrows, backgrounds}
\newcommand*\circled[1]{\tikz[baseline=(char.base)]{
            \node[shape=circle,draw, inner sep=.5pt] (char) {#1};}}
\usepackage[outline]{contour}
\contourlength{1.2pt}
\usepackage{enumitem}
\usepackage{multirow}
\usepackage{hhline}

\begin{document}

\title{Determination of the Semion Code Threshold using Neural Decoders}

\author{S.\;Varona}
\email{svarona@ucm.es}
\author{M.\;A.\;Martin-Delgado}
\email{mardel@ucm.es}
\affiliation{Departamento de F\'{\i}sica Te\'orica, Universidad Complutense, 28040 Madrid, Spain}


%\date{\today}

\begin{abstract}
We compute the error threshold for the semion code, the companion of the Kitaev toric code with the same gauge  symmetry group $\mathbb{Z}_2$. The application of statistical mechanical mapping methods is highly discouraged for the semion code, since the code is non-Pauli and non-CSS. Thus, we use machine learning methods, taking advantage of the near-optimal performance of some neural network decoders: multilayer perceptrons and convolutional neural networks (CNNs).  We find the values  $p_{\text {eff}}=9.5\%$ for uncorrelated bit-flip and phase-flip noise, and $p_{\text {eff}}=10.5\%$ for depolarizing noise. We contrast these values with a similar analysis of the Kitaev toric code on a hexagonal lattice with the same methods. For convolutional neural networks, we use the ResNet architecture, which allows us to implement very deep networks and results in better performance and scalability than the multilayer perceptron approach. We analyze and compare in detail both approaches and provide a clear argument favoring the CNN as the best suited numerical method for the semion code. 
\end{abstract}

%\keywords{}

\maketitle

\section{\label{intro}Introduction}

The robustness of quantum memories to external noise and decoherence is a key aspect along the way to fault-tolerant quantum computing.
Topological properties of quantum systems have become a resource of great importance to construct better and more robust quantum error correcting codes. The Kitaev toric code is the simplest topological code yielding a quantum memory \cite{Kitaev2003, Dennis2002}. It can be regarded as a simple two-dimensional lattice gauge theory with $\mathbb{Z}_2$ gauge group. In two dimensions, there is another lattice gauge theory with the same gauge group but different topological properties: the double semion model. The double semion model has been thoroughly studied in the search for new topological orders in strongly correlated systems, gapped, non-chiral and based on string-net mechanisms in two dimensions \cite{LevinWen2005,Freedman2004,vonKeyserlingk2013}. Although the Kitaev and the double semion models share the same gauge group, there are some remarkable differences between both. For instance, braiding two elementary quasiparticle excitations gives a $\pm 1$ phase in the Kitaev toric code, while in the double semion model it yields a $\pm \mathrm{i}$ phase factor, showing anyonic statistics. 

Topological orders can provide us with a great variety of new topological codes  with non-Pauli stabilizers \cite{fuente2020nonpauli}. These new codes might be more appropriate for practical implementation or have smaller overheads when performing, for instance, non-Clifford gates. This strongly motivates the search of new topological codes and their properties beyond the usual toric and color codes.
Recently, an error correcting code based on the double semion model, the semion code, was presented \cite{Dauphinais2019semion}. This code is topological and follows the stabilizer formalism. However, in contrast to the Kitaev toric code, it is not a CSS code \cite{PhysRevA.54.1098,PhysRevLett.77.793,RevModPhys.87.307}, since both Pauli $X$ and $Z$ operators are present in the plaquette operators, and it is not a Pauli code, since plaquette operators cannot be expressed as a tensor product of Pauli matrices. 

In order to characterize the performance and efficiency of an error correcting code, the threshold value is one of the most representative quantities quoted \cite{Dennis2002,katzgraber2009error,andrist2011tricolored,bombin2012strong}. The threshold represents the physical error rate below which increasing the distance of the code reduces the logical error rate. This error rate separates two different regimes. For error rates below threshold, larger codes translate into longer memory time and lower logical error rate. Therefore, in this regime, it makes sense to use error correction. In this work, we try to shed some light on the threshold properties of the semion code and compare it with the well-known toric code.

In the case of the Kitaev toric code, when considering Pauli noise, the threshold is determined by mapping the system to a statistical model, the random bond two-dimensional Ising model \cite{Dennis2002,katzgraber2009error,andrist2011tricolored,chubb2018statistical}. The threshold value corresponds in this new system to the phase transition between the ordered and disordered phases. Nevertheless, this mapping is extremely cumbersome in the semion code, because of the complex structure of the plaquette operators. Determining the threshold in the case where Pauli noise affects a non-Pauli code or non-Pauli noise affects a Pauli code, needs a new approach. We address this problem using machine learning and neural networks \cite{Jia_2019, chen2019machine,Poulsen_Nautrup_2019}.

Machine learning, and in particular neural networks, has been proposed in recent years as a solution for efficiently decoding stabilizer codes  \cite{Melko2017,sheth2019neural,Breuckmann2018,Varsamopoulos_2017,Varsamopoulos_2020,Baireuther2018machinelearning,Andreasson_2019,Krastanov_2017,fitzek2019deep,Liu_2019,Baireuther_2019,Chamberland_2018,ni2018neural,Nickerson_2019}. Although there are different approaches to the decoding problem using neural networks, one of the most common consists in applying a very simple decoder to the code. Afterwards the neural network tries to predict, given the syndrome measurement, the logical error produced by the simple decoder, so that this can be in turn corrected. This approach has been shown to produce near-optimal results for topological codes \cite{davaasuren2018general,maskara2018advantages}. Thus, the pseudo-threshold of these decoders should be very close to the optimal one. This makes neural networks a very suitable way of determining the threshold of a code, and specifically of the semion code.

In this paper, we use two different types of neural networks to build our decoders. First, a multilayer perceptron (MLP), a very simple feedforward neural network. Then, we present a certain type of convolutional neural network (CNN) called ResNet \cite{He2016resnet}, allowing the construction of very deep networks. Since CNNs naturally take into account the spatial structure of the code, we will see that they have multiple scalability and performance advantages in comparison to the MLP. 
The semion code is an ideal testing ground for the application of CNN methods in order to get the most out of them.
The application of CNNs is more justified in the case of the semion code than in the Kitaev \cite{Kitaev2003, Dennis2002} or color code \cite{bombin2006topological,bombin2007topological} since it allows us to take into account the complex spatial correlation of the syndrome pattern for Pauli noise. This effect is peculiar to the semion code and never studied thus far.
The results of the neural network decoders will be benchmarked against the minimal-weight perfect matching (MWPM) decoder \cite{Kolmogorov2009, edmonds1965}. While MWPM obtains very good results for the Kitaev code with independent bit-flip and phase-flip noise, it does not perform so well when plaquette and vertex syndromes are correlated, since it does not take these correlations into account. This is where neural networks will make a big difference.

The article is organized as follows. In Sec.\ \ref{sec:semion_code} we present a short review of the semion code and the noise model considered. In Sec.\ \ref{sec:neural_decoders} we introduce  the neural network decoders, compare their performance and finally present the threshold values. Sec.\ \ref{sec:conclusions} is devoted to conclusions.

%%%%%%%
%%%%%%%
\section{Error correction with the semion code} \label{sec:semion_code}

%%%%%%%%%%%%%%%%%%%%%%%%%%%%%%%%%%%%
%%%%%%%%%%%%%%%%%%%%%%%%%%%%%%%%%%%%
\begin{figure}
  \centering
  \subfloat[\label{fig:vertex_on_lattice}]{%
  \raisebox{.5cm}{
  % \documentclass{standalone}
% \usepackage{tikz}
% \begin{document}

\begin{tikzpicture}
[scale=.6]

\draw[dashed] ({cos(30)},{-sin(30)}) -- ++ (30:1) -- ++ (90:1) -- ++ (150:1) -- ++ (210:1) -- ++ (150:1) -- ++ (210:1) -- ++ (270:1) -- ++ (330:1) -- ++ (270:1) -- ++ (330:1) -- ++ (30:1) -- ++ (90:1);

% \draw[thick] (0,0) -- ++(90:.5) node[right] {1} -- ++ (90:.5);
% \draw[thick] (0,0) -- ++(-30:.5) node[above right] {3} -- ++ (-30:.5);
% \draw[thick] (0,0) -- ++(210:.5) node[above left] {2} -- ++(210:.5);

\draw[thick] (0,0) -- ++(90:.5) -- ++ (90:.5);
\draw[thick] (0,0) -- ++(-30:.5) -- ++ (-30:.5);
\draw[thick] (0,0) -- ++(210:.5) -- ++(210:.5);

\node[below] at (0,0) {$Q_{v}$};

\end{tikzpicture}

% \end{document}
%
  }}\hspace{1cm}
  \subfloat[\label{fig:plaquette_on_lattice}]{%
  % \documentclass{standalone}
% \usepackage{tikz}
% \begin{document}

\begin{tikzpicture}
[scale=.7]

\draw[dashed] ({cos(30)},{-sin(30)}) -- ++ (30:1) -- ++ (90:1) -- ++ (150:1) -- ++ (90:1) -- ++ (150:1) -- ++ (210:1) -- ++ (150:1) -- ++ (210:1) -- ++ (270:1) -- ++ (210:1) -- ++ (270:1) -- ++ (330:1) -- ++ (270:1) -- ++ (330:1) -- ++ (30:1) -- ++ (330:1) -- ++ (30:1) -- ++ (90:1);

\draw[red, thick] (0,0) coordinate (v2) -- ++(90:1) coordinate (v1) -- ++(150:1) coordinate (v6) -- ++(210:1) coordinate (v5) -- ++(270:1) coordinate (v4) -- ++(330:1) coordinate (v3) -- ++(30:1);

\draw[thick] (v1) --++ (30:1) coordinate (v1p);
\draw[thick] (v2) --++ (-30:1) coordinate (v2p);
\draw[thick] (v3) --++ (-90:1) coordinate (v3p);
\draw[thick] (v4) --++ (-150:1) coordinate (v4p);
\draw[thick] (v5) --++ (-210:1) coordinate (v5p);
\draw[thick] (v6) --++ (-270:1) coordinate (v6p);

% \draw (0,0) -- ++(90:.5) node[right] {1} -- ++ (90:.5);
% \draw (0,0) -- ++(-30:.5) node[above right] {3} -- ++ (-30:.5);
% \draw (0,0) -- ++(210:.5) node[above left] {2} -- ++(210:.5);

\node[] at ({-cos(30)}, {.5}) {$B_{p}$};

\node[] at ($(v1)!0.5!(v2)$) {\scalebox{.7}{\contour{white}{1}}};
\node[] at ($(v2)!0.5!(v3)$) {\scalebox{.7}{\contour{white}{2}}};
\node[] at ($(v3)!0.5!(v4)$) {\scalebox{.7}{\contour{white}{3}}};
\node[] at ($(v4)!0.5!(v5)$) {\scalebox{.7}{\contour{white}{4}}};
\node[] at ($(v5)!0.5!(v6)$) {\scalebox{.7}{\contour{white}{5}}};
\node[] at ($(v6)!0.5!(v1)$) {\scalebox{.7}{\contour{white}{6}}};

\node[] at ($(v1)!0.5!(v1p)$) {\scalebox{.7}{\contour{white}{12}}};
\node[] at ($(v2)!0.5!(v2p)$) {\scalebox{.7}{\contour{white}{7}}};
\node[] at ($(v3)!0.5!(v3p)$) {\scalebox{.7}{\contour{white}{8}}};
\node[] at ($(v4)!0.5!(v4p)$) {\scalebox{.7}{\contour{white}{9}}};
\node[] at ($(v5)!0.5!(v5p)$) {\scalebox{.7}{\contour{white}{10}}};
\node[] at ($(v6)!0.5!(v6p)$) {\scalebox{.7}{\contour{white}{11}}};

\end{tikzpicture}

% \end{document}
%
  }\\
  \subfloat[\label{fig:string_structure}]{
    % \documentclass{standalone}
% \usepackage{tikz}
% \begin{document}

\def\hexa#1#2#3{ 
\begin{scope}[shift={#1}, rotate=#2, scale=#3]     
    \draw[dashed] (0,0) -- ++(30:1) -- ++(90:1)
    -- ++(150:1)-- ++(210:1)-- ++(270:1)
    -- ++(330:1);
\end{scope}}

\begin{tikzpicture}
[scale=.6]

% Dashed hexagons
\foreach \j in {0,...,2}{            
\foreach \i in {0,...,1}{\hexa{({2*\j*cos(30)+\i*cos(30)},{(\i+\i*sin(30))*1})}{0}{1}} }
\hexa{({-2*cos(30)+cos(30)},{(1+sin(30))*1})}{0}{1}

\draw[thick, red] (0, {2*sin(30)+1}) coordinate (v1)-- ++(-30:1) coordinate (v2)-- ++(30:1) coordinate (v3)-- ++(-30:1) coordinate (v4)-- ++(30:1) coordinate (v5);

\draw[thick] (v1) -- ++ (90:1);
\draw[thick] (v1) -- ++ (210:1);
\draw[thick] (v2) -- ++ (-90:1);
\draw[thick] (v3) -- ++ (90:1);
\draw[thick] (v4) -- ++ (-90:1);
\draw[thick] (v5) -- ++ (90:1);
\draw[thick] (v5) -- ++ (-30:1) coordinate (v6);

\node[above] at (v4) {{\color{red} $\mathcal P$}};
\node[red] at (v1) {\textbullet};
\node[red] at (v5) {\textbullet};
\node[below right] at (v6) {$S^{\pm}_{\mathcal P}$};

\node[blue] at ($ (v5) + ({cos(30)},{0.5}) $) {\textbullet};
\node[blue] at ($ (v1) + ({-cos(30)},{0.5}) $) {\textbullet};

\end{tikzpicture}

% \end{document}
  }
  \caption{The support of each of the operators is shown by continuous lines. The qubits are placed on the edges. (a) Vertex operator $Q_{v}$. (b) Plaquette operator $B_{p}$. Note that the plaquette contains not only the red hexagon (where $X$-Pauli operators are applied), but also the outgoing legs. (c) Positive- or negative-chirality string operator $S^{\pm}_{\mathcal P}$. Path $\mathcal P$ is indicated by the solid red edges, where the $X$-Pauli operators are applied. The support of $S^{\pm}_{\mathcal P}$, $\mathrm{Conn} (\mathcal P)$, is indicated with continuous lines. The effect of $S^{\pm}_{\mathcal P}$ on the ground state of the system is to create a pair of vertex excitations at the vertices located at the endpoints of the path $\mathcal P$, which are identified by red dots. Additionally, the negative-chirality string creates a pair of plaquette excitations at the endpoint plaquettes (blue dots).}
  \label{fig:support_plaquette}
  \end{figure}
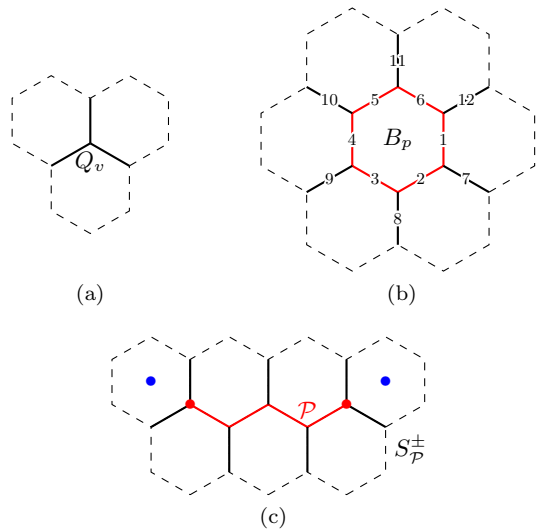
  %%%%%%%%%%%%%%%%%%%%%%%%%%%%%%%%%%%%
  %%%%%%%%%%%%%%%%%%%%%%%%%%%%%%%%%%%%

The semion code \cite{Dauphinais2019semion} is an error correcting code based on the double semion model. This error correcting code bears similarities to the Kitaev toric code. In particular, it is topological and is a stabilizer code \cite{PhysRevA.54.1862}, i.e., plaquette and vertex operators are periodically measured to detect errors. Nevertheless, the semion code is non-CSS and non-Pauli because of the structure of plaquette operators, and is defined in a hexagonal lattice. In the hexagonal lattice, edges will represent the physical qubits and vertices and plaquettes the stabilizer operators.

Vertex operators are equivalent to the ones in the Kitaev toric code $Q_{v} = Z_i Z_j Z_k$, a $Z$-Pauli operator applied on each edge of vertex $v$. Plaquettes are different; their support includes not only a hexagon, but also the outgoing legs of the hexagon (see Fig.\ \ref{fig:support_plaquette}). We have $X$-Pauli operators applied on the edges of the hexagon, as in the toric code, but we also have a diagonal operator $\sum_{\vec j} b_{p} ( \vec j ) \vert \vec j \rangle \langle \vec j \vert$ acting on the 12 qubits shown in Fig.\ \ref{fig:plaquette_on_lattice}. Thus, we have
\begin{equation} \label{eq:plaquette}
B_{p} = \prod_{k \in \partial p} X_k \ \sum_{\vec j} b_{p} ( \vec j ) \vert \vec j \rangle \langle \vec j \vert,
\end{equation}
where $b_{p} ( \vec j )$ is a function taking values in $\{\pm 1, \pm \mathrm{i}\}$, $\vec j$ is a bit string representing a state in the computational basis and $\partial p$ are the edges belonging to the border of the plaquette. $b_{p} ( \vec j )$  is given explicitly  by
\begin{equation} \label{eq:plaquette_phases}
\sum_{\vec j} b_p ( \vec j ) \vert \vec j \rangle \langle \vec j \vert =  \prod_{k \in \partial p} \left( -1 \right)^{n^{-}_{k-1} n^{+}_{k}}  \prod_{v \in p} \beta_{v},
\end{equation}
where $n^{\pm}_i = \frac{1}{2} (1 \pm Z_i)$, the subscript $v$ runs over the vertices belonging to plaquette $p$ and $\prod_{v \in p} \beta_{v}$ is
\begin{align} \label{eq:beta_algebraic}
\prod_{v \in p} \beta_{v} =&\ \mathrm{i}^{n_{12}^- \left( n^-_1 n^-_6 - n^+_1 n^+_6 \right)}\ \mathrm{i}^{n^-_7 \left( n^+_1 n^+_2 - n^-_1 n^-_2 \right)}\nonumber\\[-10pt]
\times&\ \mathrm{i}^{n^+_8 \left( n^-_2 n^+_3 - n^+_2 n^-_3 \right)}\ \mathrm{i}^{n^-_9 \left( n^-_3 n^-_4 - n^+_3 n^+_4 \right)}\\
\times&\ \mathrm{i}^{n^-_{10} \left( n^+_4 n^+_5 - n^-_4 n^-_5 \right)}\ \mathrm{i}^{n^+_{11} \left( n^-_5 n^+_6 - n^+_5 n^-_6 \right)},\nonumber
\end{align}
following the labeling of Fig.\ \ref{fig:plaquette_on_lattice}. The $X$-Pauli operators on the plaquette edges and the $(-1)$ factors of Eq.\ \eqref{eq:plaquette_phases} form the plaquette operator as defined originally in the double semion model topological order,
\begin{equation}
  \tilde{B}_p = \prod_{k \in \partial p} X_k  \prod_{k \in \partial p} \left( -1 \right)^{n^{-}_{k-1} n^{+}_{k}}.
\end{equation}
However, this operator is not Hermitian in the whole Hilbert space. Neighboring plaquettes do not commute either. The local $\beta_v$ phases added at each vertex to $\tilde{B}_p$ solve these two issues and allow us to define an error correcting code based on the stabilizer formalism. The code space of the system is formed by states with eigenvalue $+1$ for all vertex operators and $-1$ for all plaquette operators. Similarly to the Kitaev toric code, when an error occurs, the sign of some of the stabilizers flips. These locations where the stabilizer flipped can be regarded as excitations. The recovery procedure consists in annihilating the excitations with each other in such a way that the total string operator applied (the trajectory of the quasiparticles) forms a trivial loop.

String operators generating plaquette excitations, $S^Z$, are identical to the ones in the Kitaev toric code, i.e., a string of $Z$ operators. These operators commute with every stabilizer except the plaquettes at the endpoints of the string. String operators generating vertex excitations are formed by a string of $X$, as in the Kitaev code, and additionally some phases $\sum_{\vec j} F ( \vec j ) \vert \vec j \rangle \langle \vec j \vert$, where $F ( \vec j )$ takes values in $\{\pm 1, \pm \mathrm{i}\}$. For a string on a path $\cal P$ we have
\begin{equation} \label{eq:String_ansatz}
S^{+}_{\mathcal P} = \prod_{k \in \mathcal P} X_k  \sum_{\vec j } F ( \vec j ) \vert \vec j \rangle \langle \vec j \vert.
\end{equation}
The support of $S^{+}_{\mathcal P}$ is Conn($\cal P$), which is shown in Fig.\ \ref{fig:string_structure}. This means that the operator $S^{+}_{\mathcal P}$ acts non-trivially only on the set of qubits Conn($\cal P$). Thus, $F(\vec j)=F(\vec j \oplus \vec i)$ for any $\vec i$ whose qubits in Conn($\cal P$) are zero. Here, $\oplus$ denotes the sum mod 2 of the bitstrings. $F ( \vec j )$ can be determined by imposing that the string operator must square to one and that it must commute with the stabilizers (except at the endpoints, where it must anticommute). These constraints give rise to a linear system of equations from which $F ( \vec j )$ can be easily obtained. The quasiparticle vertex excitations generated by $S^{+}_{\mathcal P}$ behave like anyons. They are called semions due to the fact that their topological charge is \emph{half}  of that of a fermion, i.e., $\pm \mathrm{i}$. The negative chirality strings can be obtained multiplying $S^{+}$ by an $S^Z$ string operator joining both endpoints, $S^{-} = S^{+}S^Z$. Since $S^+$ and $S^-$ create semions at the endpoints, two strings with the same chirality crossing once anticommute, while strings with opposite chirality commute. $S^Z$ commutes with itself and anticommutes with $S^{\pm}$. Summarizing, we have, for strings crossing once, $\{S^Z, S^{\pm}\}=0$, $\{S^{\pm},S^{\pm}\}=0$ and $[S^{\pm},S^{\mp}]=0$.

\begin{figure}
  \centering
  % \documentclass{standalone}
% \usepackage{tikz}
% \begin{document}

\def\hexa#1#2#3{ 
\begin{scope}[shift={#1}, rotate=#2, scale=#3]     
    \draw (0,0) -- ++(30:1) -- ++(90:1)
    -- ++(150:1)-- ++(210:1)-- ++(270:1)
    -- ++(330:1);
\end{scope}}

\begin{tikzpicture}
[scale=1.5]

\foreach \j in {0,...,2}{            
\foreach \i in {0,...,2}{\hexa{({\j*cos(30)+\i*cos(30)/2},{(\i+\i*sin(30))*.5})}{0}{.5}} }

% Top vertical extra lines
\foreach \j in {1,...,4}{  
\draw ({2*cos(30)*.5*\j},{(sin(30)*.5+.5*3)+.5+sin(30)*.5}) -- + (90:.5);
}
% Right extra lines
\foreach \j in {0,...,3}{  
\draw ({5*cos(30)*.5+\j*cos(30)*.5},{sin(30*.5)+(.5*sin(30)+.5)*\j}) -- + (-30:.5);
}
% Left bottom extra line
\draw ({-.5*cos(30)}, {.5*sin(30)}) -- +(30:-.5);
% Right top extra line
\draw ({5*cos(30)*.5+3*cos(30)*.5},{sin(30*.5)+(.5*sin(30)+.5)*3}) -- +(30:-.5);

% Direct lattice vertical line
\draw[ultra thick, dashed, blue] (0,0) -- ++(30:.5) -- ++(90:.5) -- ++(30:.5) -- ++(90:.5) -- ++(30:.5) -- ++(90:.5) -- ++(30:.5) -- ++(90:.5) node[above] {$\bar{X}_{2} \equiv S_{{\cal V}}^+$};
% Direct lattice horizontal line
\draw[ultra thick, dashed, red] ({-.5*cos(30)},{.5*sin(30)+.5})   node[left] {$\bar{X}_{1} \equiv S_{{\cal H}}^-$} -- ++(30:.5) -- ++(-30:.5) -- ++(30:.5) -- ++(-30:.5) -- ++(30:.5) -- ++(-30:.5) -- ++(30:.5) -- ++(-30:.5);
% Reciprocal lattice horizontal line
\draw[ultra thick, dashed, blue] ({2*.5*cos(30)-.5*cos(30)},{.5*sin(30)+.5/2 + 2*sin(30)*.5 + 2*.5})  node[left] {$\bar{Z}_{2} \equiv S_{\bar{\cal H}}^Z$} -- ++ (0:{8*.5*cos(30)});
% Reciprocal lattice vertical line
\draw[ultra thick, dashed, red] ({3.5*cos(30)*.5}, {.5*.5*sin(30)}) -- ++(60:{7.65*cos(30)*.5}) node[above] {$\bar{Z}_{1} \equiv S_{\bar{\cal V}}^Z$};

% Plaquette labeling, four rows
\foreach \j in {1,...,4}{  
\node[shape=circle,draw,inner sep=1pt] at ({2*cos(30)*.5*(\j-1)},{.5*sin(30)+.5/2}) {\tiny \contour{white}{\j}};
}
\foreach \j in {5,...,8}{  
\node[shape=circle,draw,inner sep=1pt] at ({2*cos(30)*.5*(\j-5) + .5*cos(30)},{.5*sin(30)+.5/2 + sin(30)*.5 + .5}) {\tiny \contour{white}{\j}};
}
\foreach \j in {9,...,12}{  
\node[shape=circle,draw,inner sep=1pt] at ({2*cos(30)*.5*(\j-9) + 2*.5*cos(30)},{.5*sin(30)+.5/2 + 2*sin(30)*.5 + 2*.5}) {\tiny \contour{white}{\j}};
}
\foreach \j in {13,...,16}{  
\node[shape=circle,draw,inner sep=1pt] at ({2*cos(30)*.5*(\j-13) + 3*.5*cos(30)},{.5*sin(30)+.5/2 + 3*sin(30)*.5 + 3*.5}) {\tiny \contour{white}{\j}};
}

% vertex labeling, four rows
\foreach \j in {1,...,8}{   
\ifodd\j           
\node[] at ({cos(30)*.5*(\j-3)},{0}) {\tiny \contour{white}{\j}};
\else
\node[] at ({cos(30)*.5*(\j-3)},{.5*sin(30)}) {\tiny \contour{white}{\j}};
\fi}

\foreach \j in {9,...,16}{   
\ifodd\j           
\node[] at ({cos(30)*.5*(\j-10)},{.5*sin(30)+.5}) {\tiny \contour{white}{\j}};
\else
\node[] at ({cos(30)*.5*(\j-10)},{2*.5*sin(30)+.5}) {\tiny \contour{white}{\j}};
\fi}

\foreach \j in {17,...,24}{   
\ifodd\j           
\node[] at ({cos(30)*.5*(\j-19)+2*.5*cos(30)},{2*.5+2*.5*sin(30)}) {\tiny \contour{white}{\j}};
\else
\node[] at ({cos(30)*.5*(\j-19)+2*.5*cos(30)},{.5*sin(30)+2*.5+2*.5*sin(30)}) {\tiny \contour{white}{\j}};
\fi}

\foreach \j in {25,...,32}{   
\ifodd\j           
\node[] at ({cos(30)*.5*(\j-26)+2*.5*cos(30)},{.5*sin(30)+.5+2*.5+2*.5*sin(30)}) {\tiny \contour{white}{\j}};
\else
\node[] at ({cos(30)*.5*(\j-26)+2*.5*cos(30)},{2*.5*sin(30)+.5+2*.5+2*.5*sin(30)}) {\tiny \contour{white}{\j}};
\fi}

\end{tikzpicture}

% \end{document}
  \caption{Semion code embedded on a torus. Top and bottom, and left and right borders are identified. Two pairs of logical string operators (non-trivial loops going around the system) on the torus are shown (red and blue). We have 16 plaquettes, 32 vertices, and 48 physical qubits, resulting into 2 logical qubits (since one vertex and one plaquette operator are not independent). The code has distance 4. \label{fig:4_4_torus}}
\end{figure}
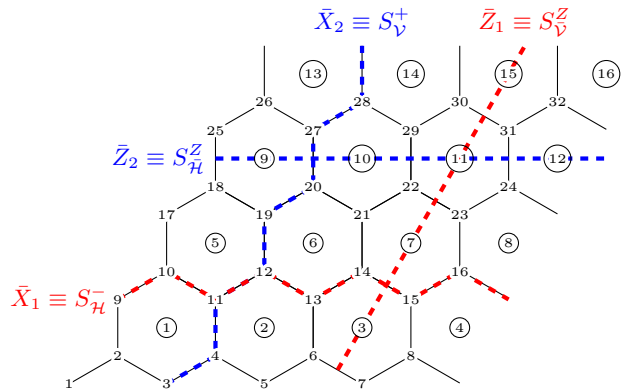

Similarly to what is done in the Kitaev code, we can embed the double semion on a torus to obtain a quantum memory with two logical qubits. An example of this can be seen in Fig.\ \ref{fig:4_4_torus}. We have a lattice with 16 plaquettes embedded on a torus. Since we have two encoded logical qubits, we need two pairs of logical operators. We can define for one of the logical qubits  $\bar{X}_{1} \equiv S_{{\cal H}}^-$ and $\bar{Z}_{1} \equiv S_{\bar{\cal V}}^Z$, and for the other $\bar{X}_{2} \equiv S_{{\cal V}}^+$ and $\bar{Z}_{2} \equiv S_{\bar{\cal H}}^Z$.
The subscript ${\cal H}$ stands for a horizontal path and ${\cal V}$ for a vertical one in Fig.\ \ref{fig:4_4_torus}.
 It is clear from the commutation rules shown previously that these set of operators fulfill the necessary anticommutation relations of the Pauli algebra. Note that the distance of $\bar{Z}$ operators is half of the $\bar{X}$, as a consequence of the hexagonal lattice. Therefore, we may be better protected against certain types of errors than against others \cite{Dauphinais2019semion,bombin2012strong,PhysRevLett.120.050505}.
To perform error correction, the stabilizers have to be measured periodically, and the excitations have to be annihilated by bringing them together using the string operators.

\subsection{Noise model}\label{sec:noise_model}
We consider Pauli noise models \cite{flammia2019efficient}, given by the expression
\begin{equation} \label{eq:noise}
\rho \rightarrow (1-p) \rho + p_X X \rho X + p_Y Y \rho Y + p_Z Z \rho Z,
\end{equation}
where $p=p_X+p_Y+p_Z$. In particular we use two error models: 
\begin{enumerate}[label=(\emph{\roman*})]
\item Independent bit-flip and phase errors with $p_X=p_Z=p_0-p_0^2$ and $p_Y=p_0^2$. Each qubit is independently acted on by an $X$ error with probability $p_0$ and by a $Z$ error with same probability $p_0$. The probability of some error happening is $p_{\rm eff}=1-(1-p_0)(1-p_0)=2p_0-p_0^2$. \label{uncorrelated_noise}
\item Depolarizing noise with $p_X=p_Y=p_Z=p_{\rm eff}/3$. With probability $p_{\rm eff}$ an error occurs in a given qubit. Each error type, $X$, $Y$ and $Z$, is equally likely.
\end{enumerate}
In order to compare the threshold values obtained for both independent and depolarizing noise, we use $p_{\rm eff}$, defined as the probability of any error occurring on a given qubit.

For these noise models consisting of Pauli operators, it will be important to determine the effect of strings of $X$-Pauli operators on a path $\cal{P}$ acting  on the code. We may rewrite a string of $X$ as
\begin{equation}
  X_{\mathcal P} = \prod_{k\in \cal{P}} X_k = S^+_{\mathcal P} \sum_{\vec j} \lbrack F_{\mathcal P} ( \vec j)  \rbrack^{\ast} \vert \vec j \rangle \langle \vec j \vert.
\end{equation}
The diagonal part can be expressed as a sum of strings of $Z$-Pauli operators,
\begin{equation}
  \sum_{\vec j} \lbrack F_{\mathcal P} ( \vec j)  \rbrack^{\ast} \vert \vec j \rangle \langle \vec j \vert = \sum_{\mathcal{Q} \in {\mathrm{Conn} (\mathcal P)}} c ( Z_{\mathcal Q} ) Z_{\mathcal Q}.
\end{equation}
Here $Z_{\mathcal Q} = \prod_{j\in \mathcal Q } Z_j$ is the multiplication of $Z$-Pauli operators acting on the set of qubits $\mathcal Q$, which are contained in $\mathrm{Conn} (\mathcal P)$. The coefficients $c(Z_{\mathcal Q} )$ are given by $c(Z_{\mathcal Q} ) = \frac{1}{2^n} \mathrm{Tr} \left( Z_{\mathcal Q}    \sum_{\vec j} \lbrack F_{\mathcal P} ( \vec j)  \rbrack^{\ast} \vert \vec j \rangle \langle \vec j \vert \right)$, with $n=|\mathrm{Conn} (\mathcal P)|$. 

Now, if we apply $X_{\mathcal P}$ to a state in the code space $|L,C\rangle$, with $L$ labeling the logical subspace and $C$ representing the eigenvalues of the stabilizers ($+1$ for vertex operators and $-1$ for plaquette operators), we obtain
\begin{equation} \label{eq:X_on_codespace}
X_{\cal{P}} \vert L,C \rangle = S^+_{\cal{P}} \sum_{\mathcal{Q} \in \mathrm{Conn}(\cal P)} c(Z_{\mathcal Q} ) Z_{\mathcal Q}  \vert L,C \rangle.
\end{equation}
$S^+_{\cal{P}}$ flips the vertices at the endpoints of $\cal{P}$, and $Z_{\mathcal Q}$ flips plaquettes at the endpoints of $\cal{Q}$, i.e., those plaquettes $p$ where only one $Z$ operator acts on $\partial p$. When the stabilizer operators are measured, the state in Eq.\ \eqref{eq:X_on_codespace} collapses. The only terms remaining in the sum are those where the $Z_{\mathcal Q}$ operator is compatible with the plaquette syndrome measured. This means $Z_{\mathcal Q}$ needs to satisfy, for every plaquette involved,  $[Z_{\mathcal Q}, B_p]_{\mathrm{s}(p)}=0$, where $\mathrm{s}(p) \in \left\lbrace\pm 1 \right\rbrace$ is the syndrome of plaquette $p$, and $[\cdot\hspace{.05cm} , \cdot]_-$ denotes the commutator and $[\cdot \hspace{.05cm}, \cdot]_+$ the anticommutator. Therefore, we have
\begin{equation} \label{psi_measure}
  \vert L,C' \rangle = N S^+_{\cal{P}} \sum_{\mathcal{Q}\in G} c(Z_{\mathcal Q}) Z_{\mathcal Q} \vert L,C \rangle,
\end{equation}
where $N$ is some normalization factor and $G=\{\mathcal{Q} \in \mathrm{Conn}(\mathcal{P} ): [Z_{\mathcal Q}, B_p]_{\mathrm{s}(p)}=0\ \forall p \in \mathcal{B}_{\mathcal{P}}\}$. $\mathcal{B}_{\mathcal{P}}$ represents the set of plaquettes whose support contains some part of $\mathcal P$. $\vert L,C' \rangle$ is a state in which the plaquettes at the endpoints of $Z_{\mathcal Q}$ are violated, as well as the vertices at the endpoints of  $S^+_{\cal{P}}$. To get back to the previous state, we need a recovery operation, $Z_{\cal{R}}$, that brings us back from $Z_{\mathcal Q} \vert L,C \rangle$ to $ \vert L,C \rangle$, where the multiplication $Z_{\cal{R}} Z_{\mathcal Q}$ forms a trivial loop of $Z$-Pauli operators in the dual lattice.
Therefore, $Z_{\cal{R}}$ corrects plaquette errors. Additionally, applying some $S^+_{\cal{O}}$ (with $\cal{O}+\cal{P}$, the symmetric difference of $\cal O$ and $\cal P$, a trivial loop), we recover the initial state $\vert L,C \rangle$.

The probability of obtaining a certain plaquette syndrome when measuring the stabilizers in state \eqref{eq:X_on_codespace} is
\begin{equation} \label{eq:probability_plaquette_syndrome}
  P(\mathrm s) = \left| \sum_{\mathcal{Q}\in G} c(Z_\mathcal{Q}) \right|^2.
\end{equation}
The different probabilities of plaquette excitations in the case of a single $X$ operator acting on a qubit can be seen in Tab.\ \ref{tab:X_effect}. Note that the probabilities depend on the orientation of the edge, showing that the code has some anisotropy.

\begin{figure}
  \subfloat[\label{fig:orientationa}]{
  % \documentclass{standalone}
% \usepackage{tikz}
% \begin{document}

\begin{tikzpicture}
[scale=.9]

\draw[dashed] (0,0)-- ++ (-90:1);
\draw[dashed] (0,0)-- ++ (150:1);
\draw[] (0,0)-- ++ (30:1) coordinate (v2);
\draw[dashed] (v2) -- ++ (90:1);
\draw[dashed] (v2) -- ++ (-30:1);

\node[] at ({-cos(30)}, {-1/2}) {$ p$};
\node[] at ({0}, {1/2+sin(30)}) {$ q$};
\node[] at ({cos(30)}, {-1/2}) {$ r$};
\node[] at ({2*cos(30)}, {1/2+sin(30)}) {$ s$};

\end{tikzpicture}

% \end{document}
  } 
  \subfloat[\label{fig:orientationb}]{
  % \documentclass{standalone}
% \usepackage{tikz}
% \begin{document}

\begin{tikzpicture}
[scale=.9]

\draw[] (0,0)-- ++ (90:1) coordinate (v2);
\draw[dashed] (0,0)-- ++ (-30:1);
\draw[dashed] (0,0)-- ++ (210:1);
\draw[dashed] (v2) -- ++ (30:1);
\draw[dashed] (v2) -- ++ (150:1);

\node[] at ({-cos(30)}, {1/2}) {$ q$};
\node[] at ({0}, {-1/2-sin(30)}) {$ p$};
\node[] at ({cos(30)}, {1/2}) {$ r$};
\node[] at (0, 2) {$ s$};

\end{tikzpicture}

% \end{document}
  }
  \subfloat[\label{fig:orientationc}]{
  % \documentclass{standalone}
% \usepackage{tikz}
% \begin{document}

\begin{tikzpicture}
[scale=.9]

\draw[dashed] (0,0)-- ++ (-90:1);
\draw[dashed] (0,0)-- ++ (30:1);
\draw[] (0,0)-- ++ (150:1) coordinate (v2);
\draw[dashed] (v2) -- ++ (90:1);
\draw[dashed] (v2) -- ++ (210:1);

\node[] at ({-2*cos(30)}, {1/2+sin(30)}) {$ p$};
\node[] at ({-cos(30)}, {-1/2}) {$ q$};
\node[] at ({cos(30)}, {-1/2}) {$ s$};
\node[] at (0, {1/2+sin(30)}) {$ r$};

\end{tikzpicture}

% \end{document}
  }
  \caption{The three possible edge orientations on which the $X$ operator can be applied. The $X$ operator acts on the central continuous edge, and may leave plaquette excitations on the four surrounding plaquettes labeled by $p$, $q$, $r$ and $s$. The probabilities of measuring a given plaquette pattern are given in Tab.~\ref{tab:X_effect}.}
  \label{fig:X_effect} 
\end{figure}
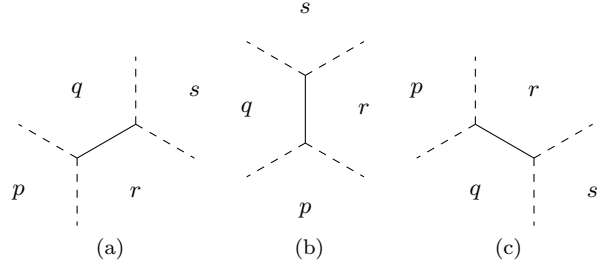

\begin{table}
  \centering
  \begin{tabular}{| c | c | c | c |}
  \cline{2-4}
  \multicolumn{1}{c|}{}
  & \multicolumn{3}{c |}{Probability} \\ \hline
  $\mathrm{s}( {p,q,r,s} )$ & \footnotesize{Orientation} (a) & \footnotesize{Orientation} (b) & \footnotesize{Orientation} (c) \\ \hline
  $\left( ---- \right)$ & $ 9/16 $ & $1/16$ & $9/16$ \\ \hline
  $\left( ++-- \right)$ & $ 1/16 $ & $ 1/16$ & $ 1/16$ \\ \hline 
  $\left( +-+- \right)$ & $1/16$ & $ 1/16$ & $ 1/16$ \\ \hline
  $\left( -++- \right)$ & $1/16$ & $9/16$ & $ 1/16$ \\ \hline
  $\left( +--+ \right)$ & $1/16$ & $ 1/16$ & $ 1/16$ \\ \hline
  $\left( -+-+ \right)$ & $1/16$ & $ 1/16$ & $ 1/16$ \\ \hline
  $\left( --++ \right)$ & $1/16$ & $ 1/16$ & $ 1/16$ \\ \hline
  $\left( ++++ \right)$ & $1/16$ & $ 1/16$ & $ 1/16$ \\ \hline
  \end{tabular}
  \caption{The various probabilities of getting a given plaquette syndrome pattern after the application of the operator $X$ on a qubit, for the three possible edge orientations \cite{Dauphinais2019semion}. The plaquettes label correspond to the ones in Fig.~\ref{fig:X_effect}. The $+$ sign represents excitations at a given plaquette.}
  \label{tab:X_effect}
  \end{table}

Given the complex structure of plaquette operators and the stabilizer syndromes caused by $X$- and $Y$-Pauli operators, the error syndromes of plaquette and vertex operators will be highly correlated for Pauli noise models, even for the case of \ref{uncorrelated_noise} independent bit-flip and phase errors. Mapping the system to a
tractable statical mechanical model in the same way it is done with the Kitaev toric code to determine the threshold is extremely difficult. This calls for alternative methods to address this problem and machine learning with neural decoders has remained unexplored for the semion code.

%%%%%%%
%%%%%%%

\section{Neural network decoders} \label{sec:neural_decoders}
One of the standard practices in neural network decoders is to train a neural network to correct the output of a simple decoder. The simple decoder is given the syndrome measurements and yields a rudimentary correction. When the correction is applied, four different outcomes may occur: the errors are corrected and the code returns to its original state ($\bar I$, identity is applied), or a logical error occurs ($\bar X$, $\bar Y$ or $\bar Z$ logical error). The neural network is trained to predict this final outcome, i.e., the logical Pauli operator applied to the code, so that the simple decoder can be corrected. In this way, the decoding process turns into a classification problem where a neural network can be used. This is the approach we will adopt here.

In particular, we consider the semion code embedded on a torus, such as the one shown in Fig.\ \ref{fig:4_4_torus}. Our simple decoder will take all excitations to the same point of the lattice, vertex number 1 or plaquette number 1, using the shortest path, to annihilate all excitations. This recovery operation produces a logical error ($\bar I$, $\bar X$, $\bar Y$ or $\bar Z$) which a neural network will try to correct. Note that since the code is embedded on a torus, we have two logical qubits. Thus, we have a total of 16 possible error combinations. Therefore, the input of our neural network will be the syndrome measurements, and the output one of these 16 categories. The input will be given as a (1d or 2d) array of bits, with value 1 corresponding to a stabilizer excitation and 0 to no excitation.

Note that one vertex and one plaquette syndrome are redundant, since they can be obtained if we know the rest of the syndromes and error excitations are created in pairs. However, in the presence of measurement errors (when the measurement of syndromes is no longer perfect) this is no longer true, and all syndromes become relevant. While our setup does not consider measurement errors and we could thus omit these two inputs, we have decided to keep them to have a more generalizable model and preserve the spatial structure of the 2d array fed into the CNN.

Training data is generated taking samples of Pauli errors according to the corresponding probability distribution of the noise model and the probability distribution of plaquette syndromes of Eq.\ \eqref{eq:probability_plaquette_syndrome}. The syndrome data is labeled with the logical error produced by the simple decoder. In the training process, the neural network is first trained on a small training set with a low error rate. Then, the network is trained with an error rate near the threshold value to obtain optimal performance. Since the error threshold is not known \textit{a priori}, several error rates are checked. A lower bound can be easily obtained by first using a MWPM decoder. Despite the fact that the model is trained for a certain error rate, it also performs well for lower error rates.

We now present two different neural network decoders. One is based on the MLP and the other is a CNN, in particular a ResNet model.

%%%%%%%
%%%%%%%

\subsection{Multilayer perceptron decoder}

\begin{figure}
  \centering
  \tikzset{%
  every neuron/.style={
    circle,
    draw,
    minimum size=.25cm
  },
  neuron missing/.style={
    draw=none, 
    scale=1,
    fill=none,
    text height=0.333cm,
    execute at begin node=\color{black}$\vdots$
  },
  hidden neuron/.style={
    circle,
    fill=gray,
    draw,
    minimum size=.25cm
  },
}

\begin{tikzpicture}[x=1.5cm, y=1.25cm, >=stealth, scale=.4]

\foreach \m/\l [count=\y] in {1,2,3,missing,4}
  \node [every neuron/.try, neuron \m/.try] (input-\m) at (0,2.5-\y) {};

\foreach \m [count=\y] in {1,2,3,4,5,missing,6}
  \node [hidden neuron/.try, neuron \m/.try ] (hidden-\m) at (2,3.75-\y) {};

\foreach \m [count=\y] in {1,2,3,4,5,missing,6}
  \node [hidden neuron/.try, neuron \m/.try ] (hidden2-\m) at (4,3.75-\y) {};

\foreach \m [count=\y] in {1,2,3,missing,4}
  \node [every neuron/.try, neuron \m/.try ] (output-\m) at (6,2.5-\y) {};

\foreach \l [count=\i] in {1,2,3,n}
  \draw [<-] (input-\i) -- ++(-1,0)
    node [left] {$s_\l$};

% \foreach \l [count=\i] in {1,n}
%   \node [above] at (hidden-\i.north) {$H_\l$};

\foreach \l [count=\i] in {\bar I \bar I, \bar I \bar X,\bar I \bar Y,\bar Z \bar  Z}
  \draw [->] (output-\i) -- ++(1,0)
    node [right] {$\l$};

\foreach \i in {1,...,4}
  \foreach \j in {1,...,6}
    \draw [gray] (input-\i) -- (hidden-\j);

\foreach \i in {1,...,6}
\foreach \j in {1,...,6}
      \draw [gray] (hidden-\i) -- (hidden2-\j);

\foreach \i in {1,...,6}
  \foreach \j in {1,...,4}
    \draw [gray] (hidden2-\i) -- (output-\j);

% \foreach \l [count=\x from 0] in {Input, Hidden, Ouput}
%   \node [align=center, above] at (\x*2,5) {\l \\ layer};

\node [align=center, above] at (0,2.4) {Input \\ layer};
\node [align=center, above] at (3, 4) {Hidden \\ layers};
\node [align=center, above] at (6, 2.4) {Output \\ layer};
% \node [align=center, above] at (\x*2,5) {\l \\ layer};

\end{tikzpicture}
  \caption{The MLP receives the syndrome as input, i.e., a bit string with the vertex and plaquette operator measurements, and outputs the predicted error of the two logical qubits encoded in the torus.\label{fig:mlp}}
\end{figure}
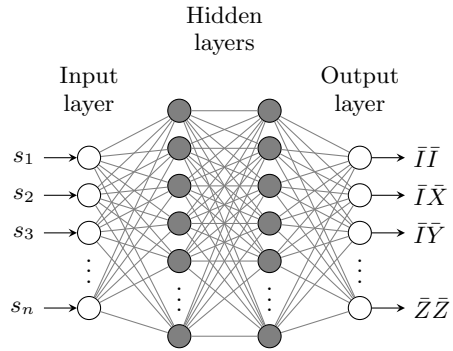

\begin{figure}
  \centering
  \includegraphics{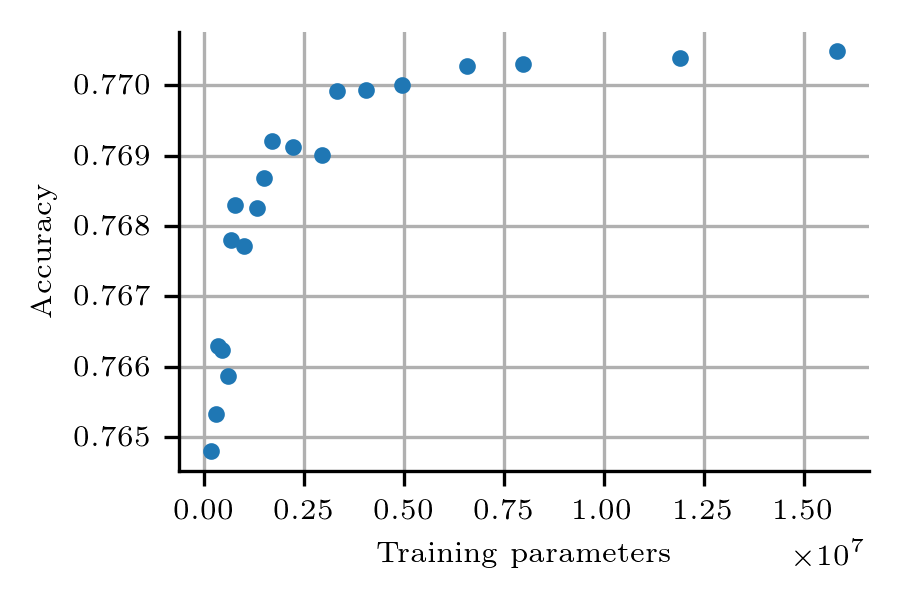}
  \caption{Accuracy as a function of the number of training parameters for MLPs with different numbers of layers and nodes. Twenty different MLPs considered with the following parameters: hidden layers, $H\in\{4,6,8,10\}$; nodes per layer, $N\in\{266, 400, 600, 900, 1400\}$. Independent bit-flip and phase error at rate $p_0=0.045$ for code distance 5.\label{fig:parameters_scaling}}
\end{figure}

The MLP is  one of the simplest classes of feedforward artificial neural networks. A MLP consists of three different parts: an input layer, hidden layers and an output layer, as can be seen in Fig.\ \ref{fig:mlp}. The layers are formed by neurons or nodes, with trainable parameters. Each node is fully connected to all nodes in the neighboring layers. The universal approximation theorem \cite{leshno1993,HORNIK1991} states that a finite MLP can approximate any continuous function. Therefore, if trained appropriately, the MLP should be a near-optimal decoder.

The structure of this MLP follows closely the one presented in Ref.\ \cite{maskara2018advantages}, where they found near-optimal decoders for other topological codes, the Kitaev code an the color code. All hidden layers have the same number of nodes. The cost function is chosen to be categorical cross entropy \cite{Goodfellow2016} and the optimizer is Adam \cite{kingma2014adam}, a gradient-based optimization algorithm with better performance than a simple gradient descent. The activation function is ReLU, $f(x)=\mathrm{max}(0,x)$, with He initialization of weights \cite{He2015}. In order to train deep neural networks and avoid vanishing gradient convergence problems, we make use of batch normalization \cite{ioffe2015batch} in each layer.

Instead of computing the gradient of the cost function in the whole dataset, an approximation is computed using a small batch of data and then the parameters are updated. The batch size was chosen to be $10^4$. The final performance of the MLP is not affected by this number, provided it is not too small. If we have a small dataset, and the network is trained several epochs over the same data, it is likely we will suffer overfitting. To avoid this, each batch of data is fed only once into the network during the training process, although this requires larger training sets.

Regarding hyperparameter tuning, a search was done to obtain the optimal number of layers and nodes. In general, for a given code distance, the higher the number of trainable parameters in the model, the better the performance, as can be seen in Fig.\ \ref{fig:parameters_scaling}. However, we found that there is a point beyond which increasing the number of layers or nodes (and consequently increasing the number of parameters) produces very little accuracy improvements while increasing substantially the training time. Beyond that point, there is a broad range of models with very similar performance and different hyperparameters. Among them, we tried to choose the one with the lowest training time. Increasing the code distance by one was roughly observed to require double as many trainable parameters in the MLP model to reach good performance.

%%%%%%%
%%%%%%%

\subsection{Convolutional neural network decoder}

%%%%%%%%%%%%%%%%%%%%%%%%%%%%%%%%%%%%
%%%%%%%%%%%%%%%%%%%%%%%%%%%%%%%%%%%%
\begin{figure}
  \centering
  \subfloat[\label{fig:resnet_block}]{%
  \raisebox{.5cm}{
  \begin{tikzpicture}[scale=.9, every node/.style={scale=.8}]
    \node[above] (x) at (0,.75) {$x$};
    \node[draw,thick] (rect1) at (0,0)  {Convolutional layer};
    \node[draw,thick] (rect2) at (0,-1)  {Convolutional layer};
    \draw [thick,->] (0,.75)  -- (rect1);
    \draw [thick,->] (rect1) -- (rect2);
    \node[right] at (0,-.5) {ReLU};

    \node[draw, thick, circle] (add) at (0,-2) {+};
    \draw [thick,->] (rect2) -- (add);
    \draw [thick,->] (add) -- (0,-2.75);

    % \draw [thick,->] plot [smooth] coordinates {(0,0.6) (1.6,.4) (2, -.75) (1.6,-1.85) (.33,-2)};
    \draw [thick,->, rounded corners] (0,0.6) -- (1.8,.6) -- (1.8,-2) -- (.28,-2);
    \draw [dashed, rounded corners] (0,0.4) -- (1.5,.4) -- (1.5,-1.4) -- (-1.5,-1.4) -- (-1.5,.4) -- cycle;
    % \node[right] at (2, -.75) {$x$};
    \node at (-2, -.5) {$f(x)$};
    \node[right] at (0,-2.5) {ReLU};
    \node[left] at (-.5,-2) {$f(x)+x$};

\end{tikzpicture}%
  }}\hspace{.8cm}
  \subfloat[\label{fig:resnet_scheme}]{%
  \begin{tikzpicture}[scale=.9, every node/.style={scale=.8}]
\node[above] (x) at (0,.75) {input};
\node[draw,thick,align=center] (rect1) at (0,0)  {$3\times 3$, 16, convolution};
\node[draw,thick] (rect2) at (0,-1)  {[$3\times 3$, 16, building block]$\times n$};
\node[draw,thick] (rect3) at (0,-2)  {[$3\times 3$, 32, building block]$\times n$};
\node[draw,thick] (rect4) at (0,-3)  {[$3\times 3$, 64, building block]$\times n$};
\node[draw,thick] (rect5) at (0,-4)  {dense layer};
\node[right] at (0,-3.5) {flatten};

\draw [thick,->] (0,.75)  -- (rect1);
\draw [thick,->] (rect1) -- (rect2);
\draw [thick,->] (rect2) -- (rect3);
\draw [thick,->] (rect3) -- (rect4);
\draw [thick,->] (rect4) -- (rect5);
% \draw [thick,->] (rect5) -- (0,-4.75);

% \node[right] at (1.7,-1) {$\times n$};
% \node[right] at (1.7,-2) {$\times n$};
% \node[right] at (1.7,-3) {$\times n$};

\end{tikzpicture}%
  }
  \caption{(a) ResNet building block. A shortcut connection skips the convolutional layers. (b) ResNet model. $n$ building blocks are stacked at each stage. The first stage has 16 filters, the second has 32 and the third 64. The filter size is always $3\times 3$ and stride equals 1. The depth of the model is $d = 6n+2$.}
  \label{fig:resnet_architecture}
  \end{figure}
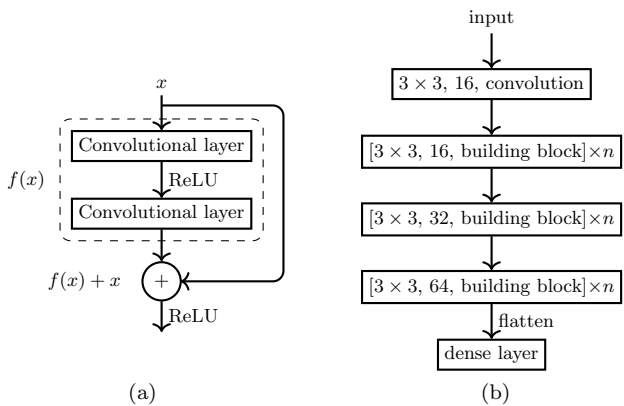
  %%%%%%%%%%%%%%%%%%%%%%%%%%%%%%%%%%%%
  %%%%%%%%%%%%%%%%%%%%%%%%%%%%%%%%%%%%

  \begin{table}
    \centering
    \begin{tabular}{c | cccccccc | c}
  
      4 & 5&6&7&8&1&2&3&4 &5 \\ \hline
  
      \circled{14}  &  \large$\times$&\circled{15}&\large$\times$&\circled{16}&\large$\times$&\circled{13}&\large$\times$&\circled{14}  &  \large$\times$ \\ 
  
      29 & 30&31&32&25&26&27&28&29 &30 \\
  
      \large$\times$  &  \circled{11}&\large$\times$&\circled{12}&\large$\times$&\circled{9}&\large$\times$&\circled{10}&\large$\times$  &  \circled{11}\\
  
      22 & 23&24&17&18&19&20&21&22 &23 \\
  
      \circled{7}  &  \large$\times$&\circled{8}&\large$\times$&\circled{5}&\large$\times$&\circled{6}&\large$\times$&\circled{7}  &  \large$\times$ \\ 
  
      15 & 16&9&10&11&12&13&14&15 &16 \\
  
     \large$\times$  &  \circled{4}&\large$\times$&\circled{1}&\large$\times$&\circled{2}&\large$\times$&\circled{3}&\large$\times$  &  \circled{4}\\
  
      8 & 1&2&3&4&5&6&7&8 &1 \\ \hline
  
      \circled{16}  &  \large$\times$&\circled{13}&\large$\times$&\circled{14}&\large$\times$&\circled{15}&\large$\times$&\circled{16}  &  \large$\times$
  
    \end{tabular}
    \caption{Mapping of syndromes of the hexagonal lattice in Fig.\ \ref{fig:4_4_torus} to a square structure suitable for a CNN. The data inside the square contains the stabilizers in the torus of Fig.\ \ref{fig:4_4_torus}, while the data outside is the periodic padding used in each convolution, indicating the periodic boundary conditions of the torus. $\times$ represents some extra values needed to preserve the hexagonal spatial structure in the square lattice; they will always be set to zero. Circled numbers represent plaquettes; the rest are vertices. The 48 stabilizer measurements of a distance $d=4$ code are fed into the CNN as a $8\times 8$ image. In general a code of distance $d$ will produce an image of size $2d\times2d$.}
    \label{tab:data2img}
  \end{table}

Despite the very good performance of the MLP in terms of accuracy, this approach is not scalable for error correction in large codes, since the training time increases exponentially with the size of the code. In addition, the information about the spatial distribution of the syndromes was not provided to the MLP, which has to figure this out by itself during training. These two problems suggest using a CNN for the task. In a CNN, the hidden layers are convolutional layers. Each input to the next layer is computed from a small local region of the preceding layer using some trainable parameters called filters. CNNs have been extensively used for image recognition. Although the information at each position in our lattice is binary, namely, the $\pm 1$ value of the stabilizer measurement (compare this to an RGB image with 256 values per pixel in each channel), we can still see it as an image and feed it into a CNN.
A syndrome pattern of errors can be considered as an image to be recognized with machine learning CNN methods.
The semion code is very special since the structure of plaquette stabilizers causes complex correlations between $X$ and $Z$ errors produced externally (see Eq.\ \eqref{eq:probability_plaquette_syndrome}). This is why we may argue that CNN decoders are especially well suited to the semion code in comparison to other neural network decoders, for CNN models were devised to mitigate the drawbacks posed by the MLP architecture by exploiting the strong spatially local correlation present in natural images.

We base our CNN model on the ResNet architecture. The ResNet architecture allows us to build very deep models, stacking a large number of convolutional layers without learning degradation. This is made possible by introducing residual shortcuts, connections performing the identity mapping and skipping the stacked layers. The shortcut output is added to the output of the stacked layers, as shown in Fig.\ \ref{fig:resnet_block}, which constitutes the building block of ResNet. Our architecture consists of three stacked stages, where each stage has $n$ building blocks like the one depicted in Fig.\ \ref{fig:resnet_block}, and the convolutional layers have $3\times 3$ filters. Batch normalization is performed after each convolutional layer. When going from one stage to the next, the number of filters is doubled. When doing image classification, this doubling in the number of filters is usually accompanied by downsampling the data using a convolutional layer of stride 2. However, we found that downsampling reduces noticeably the accuracy of the model, specially when doing two of them (at the end of stage 1 and at the end of stage 2). Therefore, we do not perform any downsamplings. At the beginning of stages 2 and 3, when the number of filters is doubled, the identity shortcut connection of the first block is substituted by a convolutional layer with the corresponding number of $1\times 1$ filters. Finally, the output of these three stages is flattened and fed into a fully connected layer with softmax activation function. The model is shown schematically in Fig.\ \ref{fig:resnet_scheme}.

In order to perform convolutions, we need to recast our hexagonally distributed data into a square lattice. This is done as presented in Tab.\ \ref{tab:data2img} for the code of distance 4 in Fig.\ \ref{fig:4_4_torus} (this transformation is explained in more detail in App.\ \ref{sec:mapping} for a code of arbitrary distance). With this mapping, some extra syndromes are introduced so that the spatial structure is faithful to the original one. These extra syndromes are always set to zero. It is also important to preserve the periodic boundary conditions of the torus. Therefore, before each convolution, the square data is padded periodically, as can be seen in Tab.\ \ref{tab:data2img}, where one extra row and one extra column is added at each side.

Again, the cost function is the categorical cross entropy and the optimizer is Adam. We also use the ReLU activation function and He initialization. The batch size is chosen to be 1000. For smaller sizes, it is more likely that the model converges to a local minimum during training. During training, when the loss value reaches a plateau, the learning rate is reduced by a factor of 0.3. This is repeated until we observe that reducing the learning rate does not produce any accuracy gains.

These models already have similar performance to the MLP in terms of classification accuracy with $n=2$, i.e., a depth of 14, and reduce substantially the number of parameters, the training time, and the size of the data set. The optimal performance was found for $n=8$, depth of 50, for all code distances considered.

Another advantage of CNNs is the possibility of using transfer learning. The parameters learned by the convolutional layers for a given code distance can be reused for another code distance; this applies especially to the initial layers of the model, which tend to learn generic features. These parameters can be used as the starting point of the optimization to reduce the training time of the model. In addition, some of these initial layers can be declared non-trainable so that the learning time is even shorter.

\begin{figure*}
  \subfloat[\label{fig:MWPM_uncorrelated_DS_threshold}]{%
    \includegraphics[width=.33\linewidth]{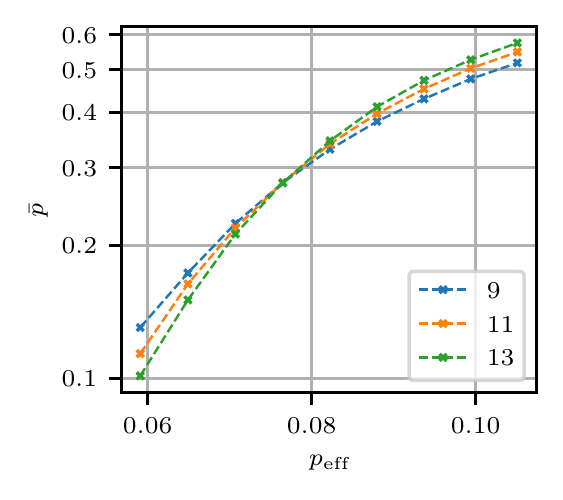}%
  }\hfill
  \subfloat[\label{fig:MLP_uncorrelated_DS_threshold}]{%
    \includegraphics[width=.33\linewidth]{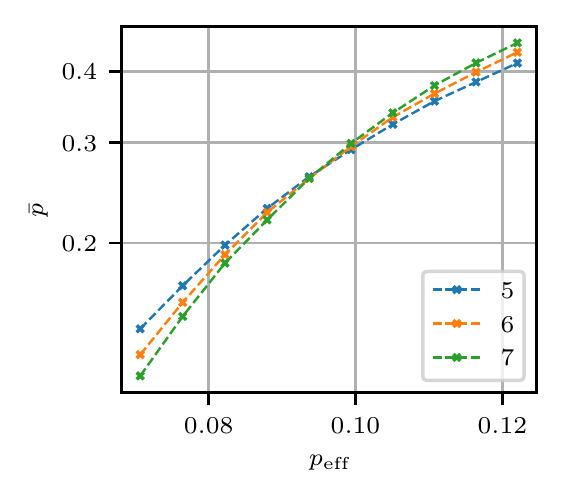}%
  }\hfill
  \subfloat[\label{fig:Resnet_uncorrelated_DS_threshold}]{%
    \includegraphics[width=.33\linewidth]{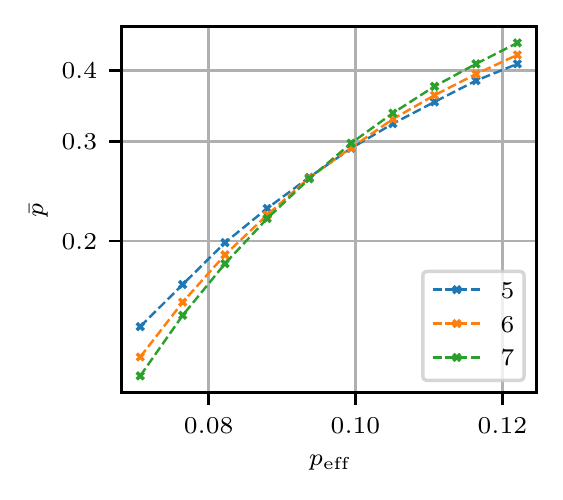}%
  }\hfill
  \subfloat[\label{fig:MWPM_depolarizing_DS_threshold}]{%
    \includegraphics[width=.33\linewidth]{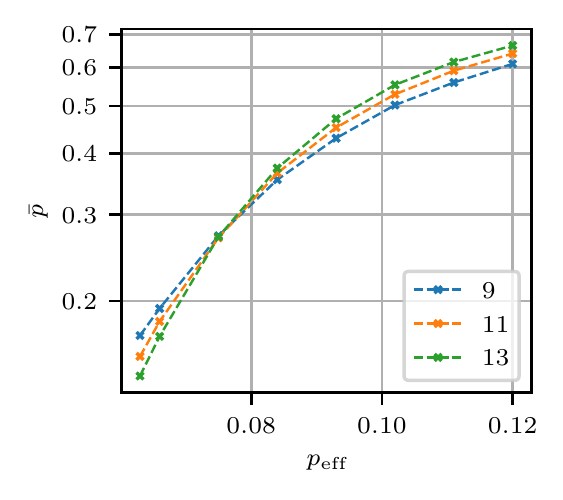}%
  }\hfill
  \subfloat[\label{fig:MLP_depolarizing_DS_threshold}]{%
    \includegraphics[width=.33\linewidth]{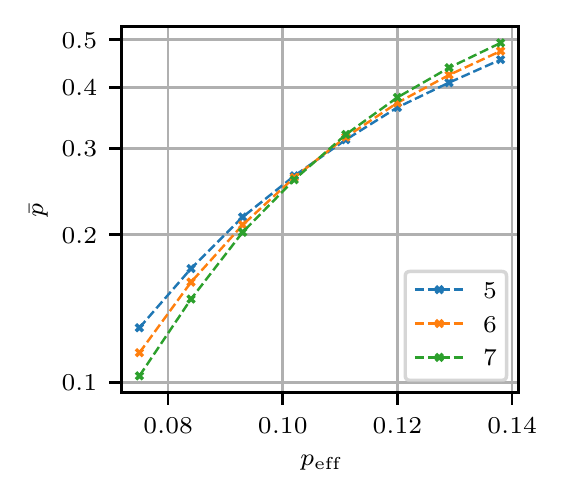}%
  }\hfill
  \subfloat[\label{fig:Resnet_depolarizing_DS_threshold}]{%
    \includegraphics[width=.33\linewidth]{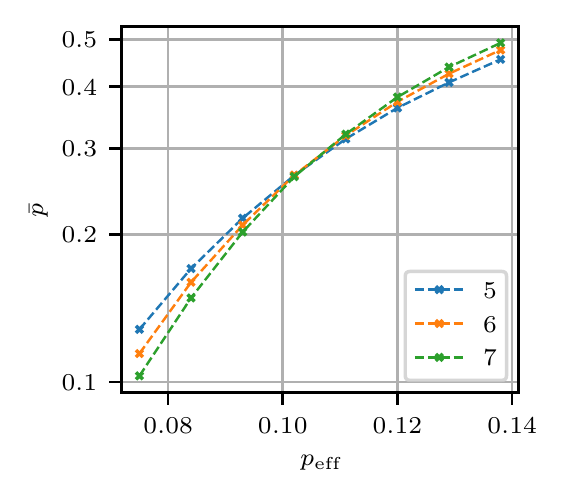}%
  }
  \caption{Logical error rate, $\bar{p}$, as a function of effective error rate, $p_{\rm eff}$. Independent bit- and phase-flip noise results in (a), (b) and (c). Depolarizing noise in (d), (e) and (f).  For the MLP, (b) and (e), the number of hidden layers, $H$, and number of neurons, $N$, are: $H=6$ and $N=900$, for distance 5; $H=7$ and $N=1100$, for distance 6; and $H=8$ and $N=1400$, for distance 7. (a) MWPM decoder with independent noise. (b) MLP decoder with independent noise. (c) ResNet50 decoder with independent noise. (d) MWPM decoder with depolarizing noise. (e) MLP decoder with depolarizing noise. (e) ResNet50 with depolarizing noise.}
  \label{fig:DS_threshold}
\end{figure*}

\subsection{Results}

Here we present the threshold values obtained for the semion code as well as an analysis of the performance of the different decoders. As was mentioned earlier, ResNet is a much better option than MLP in terms of scalability and computational costs. This is shown in Tab.\ \ref{tab:MLPvsResNet}, where we can see that the number of training steps and the number of trainable parameters are, respectively, one and two orders of magnitude lower for the ResNet model. Since the number of training steps is lower, the dataset is also smaller for ResNet (the number of training examples is obtained by multiplying the training steps by the batch size), which also reduces the cost of producing the training data.

\begin{table}[]
  \begin{tabular}{lccc}
  \cline{2-4}
  \multicolumn{1}{c}{} & T. parameters    & Steps & Accuracy \\ \hline
  MLP                  & $1.2\times 10^7$ & $6.9\times 10^5$    & 74.0\%         \\
  ResNet14             & $3.8\times 10^5$ & $9.1\times 10^4$  & 73.8\%         \\
  ResNet50             & $9.6\times 10^5$ & $4.3\times 10^4$  & 74.1 \%         \\ \hline
  \end{tabular}
  \caption{MLP and ResNet figures for code distance 7 and trained with independent noise at $p_{\rm eff}=0.048$. The MLP has $H=8$ and $N=1400$. The first column shows the number of trainable parameters of each model, the second the number of training steps, and the third the accuracy. This compares to a MWPM accuracy of 58.9\%.} \label{tab:MLPvsResNet}
\end{table}

\begin{table}[]
  \begin{tabular}{ccccc}
  \cline{3-5}
                                                   &                 & MWPM & MLP & ResNet50 \\ \hline
  \multicolumn{1}{l}{\multirow{2}{*}{SC}}          & Bit-/phase-flip & 7.6\%& 9.4\% & 9.5\% \\ 
  \multicolumn{1}{l}{}                             & Depolarizing    & 7.5\%& 10.5\% & 10.5\% \\ \hline \hline
  \multicolumn{1}{l}{\multirow{2}{*}{KTC\hspace{.2cm}}}& Bit-/phase-flip & 12.5\% &  -   & 13.2\% \\ 
  \multicolumn{1}{l}{}                             & Depolarizing    & 10.0\% &  -   & 11.9\%         \\ \hline
  \end{tabular}
  \caption{Semion code (SC) and hexagonal Kitaev toric code (KTC) $p_{\rm eff}$ threshold values for the different decoders considered: minimal weight perfect matching (MWPM), multilayer perceptron (MLP) and a ResNet50 convolutional neural network. MLP values were not computed for the KTC.}
  \label{tab:threshold}
\end{table}

In Tab.\ \ref{tab:threshold} and Fig.\ \ref{fig:DS_threshold}, we can see the threshold values for the different decoders. We obtain a threshold of 9.5\% in the case of independent bit- and phase-flip and 10.5\% for depolarizing noise. These quantities correspond to the effective error rate, $p_{\rm eff}$, defined in Sec.\ \ref{sec:noise_model}. These values contrast with the ones obtained for the 
Kitaev toric code in a hexagonal lattice, which we also obtain with neural decoders \footnote{For independent bit- and phase-flip noise, the error threshold of the Kitaev code in different lattice geometries was computed in Ref.\ \cite{Keisuke2012} using MWPM.}
(see Tab.\ \ref{tab:threshold}). 
We find the depolarizing threshold to be higher than the one for independent bit- and phase-flip noise, suggesting that plaquette and vertex syndrome correlations in the semion code play an important role. Despite the lower threshold results obtained for the semion code, it is important to note that we are considering Pauli noise. Since the stabilizers of the toric code are formed by Pauli operators, Pauli noise results in a simple structure for the errors, while for the double semion we have a much more complex structure, see Eq.\ \eqref{eq:probability_plaquette_syndrome}. This is specially so in the case of independent bit- and phase-flip noise, where plaquette and vertex syndromes are not correlated in the Kitaev toric code. We can see that the MWPM threshold gets much closer to the neural decoder threshold for the Kitaev code, since MWPM does not take into account plaquette and vertex correlations, while for the rest of the cases, where correlations contribute significantly, MWPM falls behind and neural decoders perform significantly better. As a consequence of the non-Pauli nature of the semion code, the threshold difference between depolarizing and independent noise is not as high for the semion code,
since for both noise cases vertex and plaquette syndromes are correlated.

Despite ResNet being more accurate than the MLP and intrinsically including the spatial information that the MLP lacks, thresholds obtained are nearly the same in both cases. This may suggest that the pseudo-thresholds achieved are very close to the optimal one, and there is little room for performance enhancements. 

\section{Conclusion}\label{sec:conclusions}
Quantum error correction is expected to be a fundamental tool to achieve the desired reliable and robust quantum computation and the first proof-of-principle steps towards this goal have been already achieved experimentally \cite{nigg2014quantum,muller2016iterative,barends2014,corcoles2015}.
Topological quantum error correction with Abelian stabilizer codes has become a mature research field by now providing one of the most valuable schemes on the road of fault-tolerant quantum computation \cite{RevModPhys.87.307,lidar_brun_2013,fujii2015quantum}. It all begun with the simple Kitaev toric code, whose companion model with the same gauge symmetry group -- the double semion model -- has remained outside the quantum error correction methods until recently \cite{Dauphinais2019semion}.

We have determined a near-optimal threshold for the semion code in the cases of independent bit- and phase-flip noise and depolarizing noise. The fact that neural decoders can have near-optimal performance shows that the pseudo-threshold values obtained for the semion code with the ResNet decoder should be very close to the real threshold values. Since for the semion code, the plaquette stabilizer operators are not a simple product of Pauli operators, the usual mapping to a statistical mechanical model becomes a very complex problem, and using deep learning models becomes a nice and efficient way of determining the threshold of the semion code. These same methods could be used for other topological codes with non-Pauli stabilizers or non-Pauli noise models \cite{fuente2020nonpauli,song2019twisted}.

The ResNet architecture has shown a good performance in error correction. Nevertheless, it may still be possible to obtain little performance improvements by, for instance, taking other mappings from the hexagonal lattice to the square lattice (see Tab.\ \ref{tab:data2img}) or implementing slightly different versions of the ResNet model. In addition, it would be interesting to apply this kind of deep learning models, not only to threshold determination, but also to build general purpose scalable decoders. Data augmentation, i.e., taking advantage of the symmetries of the error correcting code to reduce the size of the dataset, as was suggested in Ref.\ \cite{wagner2019symmetries}, and transfer learning, i.e., reusing the weights previously learned for smaller systems should help us to obtain scalable neural decoders. 

The source code of the neural-network decoder can be found at \url{https://github.com/varona/nn_decoder}.

%%%%%%%
%%%%%%%
\begin{figure}
  \subfloat[\label{fig:PL_vs_d_un}]{%
    \includegraphics[width=1.\linewidth]{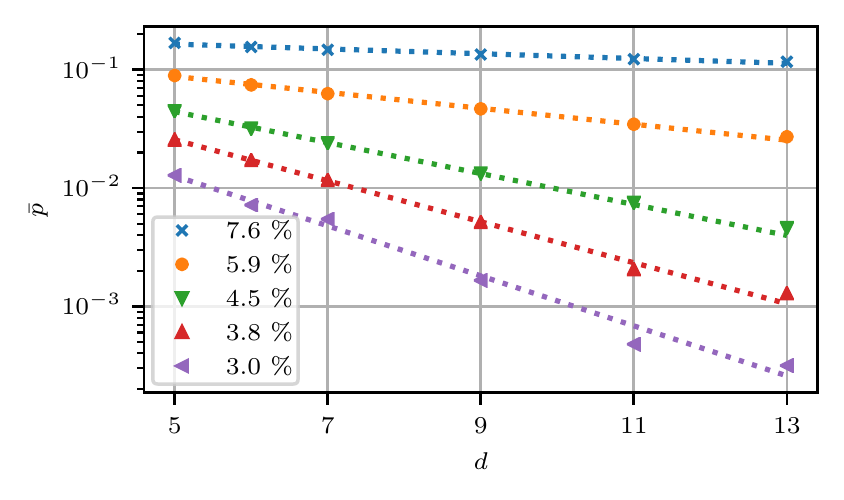}%
  }\hfill
  \subfloat[\label{fig:PL_vs_d_dep}]{%
    \includegraphics[width=1.\linewidth]{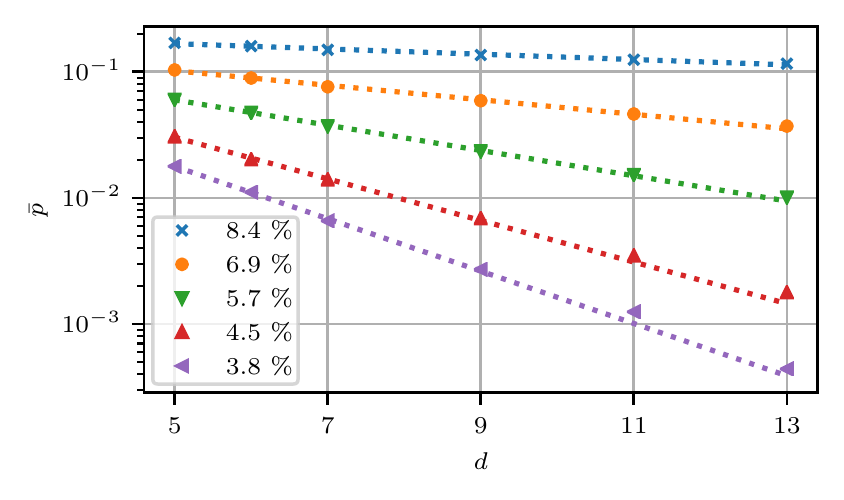}%
  }
  \caption{Logical error rate, $\bar p$, as a function of code distance, $d$, for the ResNet50 decoder. This shows the exponential suppression of noise for (a) independent bit- and phase-flip and (b) depolarizing noise. Each dotted line is an exponential fit of the data points corresponding to a $p_{\mathrm{eff}}$ value.\label{fig:exp_suppression}}
\end{figure}

%%%%%%%
%%%%%%%

\begin{acknowledgments} 
  We thank G.\ Dauphinais for useful discussions at the early stage of this research.
  The authors thankfully acknowledge the resources from the supercomputer ``Cierzo'' , HPC infrastructure of the Centro de Supercomputaci\'on de Arag\'on (CESAR), and the technical expertise and assistance provided by BIFI (Universidad de Zaragoza). S.V. especially thanks H\'ector Villarrubia Rojo for computational resources and technical assistance. We acknowledge financial support from the Spanish  MINECO  grants  MINECO/FEDER Projects FIS2017-91460-EXP and PGC2018-099169-B-I00FIS-2018 and from CAM/FEDER Project No. S2018/TCS-4342 (QUITEMAD-CM). The research of  M.A.M.-D.  has  been  partially  supported  by  the U.S.   Army   Research   Office   through   Grant   No. W911NF-14-1-0103.  S.V. thanks the support of a FPU MECD Grant.
\end{acknowledgments}

\appendix
\section{Exponential suppression of noise}

To further confirm that what we see is an error correcting threshold, we check here that ($i$) the noise is exponentially suppressed for larger code distances and ($ii$) the logical error rate is reduced to values much lower than the physical error rate. In order to show this clearly, we train ResNet50 decoders for code distances up to $d=13$. Showing this same results for the MLP is very costly given the scaling of the model. Due to memory limitations, the batch size was chosen to be 300 when training the ResNet50 models with $d>7$. These results are shown in Fig.\ \ref{fig:exp_suppression}.

\section{Mapping of the hexagonal lattice into a square lattice} \label{sec:mapping}

In this Appendix, we describe how the 1d array of stabilizer measurements provided as input to the MLP is converted into a 2d array suitable for the CNN and reflecting the spatial structure of the code. Since the semion code is defined on a hexagonal lattice, we need to convert the hexagonally distributed stabilizer measurements into a square distribution, while preserving the spatial structure. This is done as follows. The code of distance $d$ has $n_v = 2d^2$ vertices and $n_p = d^2$ plaquettes, for a total of $n= n_v+n_p=3d^2$ stabilizers. The 1d array of $n$ stabilizers measurements will be converted into a $2d\times 2d$ image, $I$. Suppose vertices and plaquettes have been sequentially labeled from left to right and bottom to top, as shown in Fig.\ \ref{fig:4_4_torus} for the code of distance 4. The syndrome of vertex $v$ corresponds to the image element $I_{i,j}$, where $i$ and $j$ are
% \begin{equation}
%     i = 2 \left\lfloor \frac{v-1}{2d} \right\rfloor +1,
% \end{equation}
% \begin{equation}
%   j = \mathrm{mod}\left(v-1 +(1-2d)\left\lfloor \frac{v-1}{2d} \right \rfloor, 2d\right)+1.
% \end{equation}
\begin{gather}
  i = 2 \left\lfloor \frac{v-1}{2d} \right\rfloor +1,\\
  j = \mathrm{mod}\left(v-1 +(1-2d)\left\lfloor \frac{v-1}{2d} \right \rfloor, 2d\right)+1.
\end{gather}
The syndrome of plaquette $p$ corresponds to the element $I_{i,j}$, with $i$ and $j$ given by
% \begin{equation}
%   i = 2\left\lfloor \frac{p-1}{d} \right\rfloor+2,
% \end{equation}
% \begin{equation}
%   j = \mathrm{mod}\left(2p + (1-4d) \left\lfloor \frac{p-1}{d} \right\rfloor , 2d\right) + 1.
% \end{equation}
\begin{gather}
    i = 2\left\lfloor \frac{p-1}{d} \right\rfloor+2, \\
    j = \mathrm{mod}\left(2p + (1-4d) \left\lfloor \frac{p-1}{d} \right\rfloor , 2d\right) + 1.
\end{gather}
Since $I$ has $4d^2$ elements and we have $3d^2$ stabilizers, there are a few elements in $I$ which do not correspond to any stabilizer. These elements are always set to zero. The result of this transformation is shown in Tab.\ \ref{tab:data2img} for the code of distance 4 of Fig.\ \ref{fig:4_4_torus}. In Tab.\ \ref{tab:data2img}, those elements denoted with the symbol ``$\times$'' represent the elements that do not correspond to any stabilizer measurement.

\bibliography{citations.bib}

\end{document}